\documentclass[12pt,preprint]{aastex}

\shorttitle{Disappearance of Penumbral Fine-Scale Structure and Evershed Flow}
\shortauthors{Kubo et al.}

\begin{document}

\title{Temporal Relation between Disappearance of Penumbral Fine-Scale
Structure and Evershed Flow} 

\author{M. Kubo}
\affil{National Astronomical Observatory, Mitaka, Tokyo, 181-8588, Japan.} 
\email{masahito.kubo@nao.ac.jp}
\author{K. Ichimoto}
\affil{Hida Observatory, Kyoto University, Takayama, Gifu 506-1314, Japan.} 
\author{B. W. Lites}
\affil{High Altitude Observatory, National Center for Atmospheric
Research, P.O. Box 3000, Boulder, CO 80307\altaffilmark{1}.}  
\author{R. A. Shine}
\affil{Lockheed Martin Solar and Astrophysics Laboratory, Building 252,
3251 Hanover Street, Palo Alto, CA 94304.}  

\altaffiltext{1}{The National Center for Atmospheric Research is sponsored by the National Science Foundation}

\begin{abstract}
We investigate the temporal relation between the Evershed flow, dot-like bright
 features (penumbral grain), the complex magnetic field structure, and
 dark lanes (dark core) along bright filaments in a sunspot penumbra.  
We use a time series of high spatial resolution photospheric intensity, vector magnetic field maps, and Doppler
 velocity maps obtained with the Solar Optical Telescope
 aboard the \textit{Hinode} spacecraft. 
We conclude that the appearance and disappearance of the Evershed flow
 and penumbra grains occur at nearly the same time and are associated with
 changes of the inclination angle of the magnetic field from vertical to more horizontal.
This supports the idea that Evershed flow is a result of thermal
 convection in the inclined field lines. 
The dark core of the bright penumbral filament also appears coincidental with 
 the Evershed flow.
However, the dark-cored bright filament survives
 at least for 10-20 minutes after the disappearance of the Evershed flow.   
The heat input into the bright filament continues after the end of heat
 transfer by the Evershed flow.  
This suggests that local heating along the bright filament is
 important to maintain the brightness of the bright filament in addition
 to the heat transfer by the Evershed flow.
\end{abstract}

\keywords{Sun: magnetic fields --- Sun: photosphere --- (Sun:) sunspots}

\section{INTRODUCTION}
Sunspot penumbrae consist of many fine-scale radial filaments.
Their fine-scale structures reflect complex magnetic field structures in
penumbrae: azimuthal fluctuations of the penumbral magnetic field
inclination and field strength are well observed at the scale of the
intensity fluctuations \citep[e.g.][]{Degenhardt1991, Schmidt1992, Title1993,
Lites1993}.
A systematic outward flow called Evershed flow is important for
understanding the formation of such complex magnetic filed structures in
penumbrae.  
Evershed flows are observed as radial filamentary structures of Doppler
blueshifts at the disk-center side penumbra, and Doppler redshifts at
the limb side penumbra. 
The relation between penumbral fine-scale structures and Evershed flows
have been investigated by numerous authors for a long period, and recent
observations with a high spatial resolution clearly confirms the 
following properties \citep[e.g.][]{Bellot2004, Langhans2005, Rimmele2006,
Ichimoto2007a}:  
(1) The Evershed flow is a nearly horizontal outward flow observed with
horizontal magnetic field component in the penumbra.
(2) The horizontal field component is observed along the bright filament
in the inner penumbra, and the dark filament in the outer penumbra.
(3) The Evershed flow originates at a dot-like bright feature called a
penumbral grain in the inner penumbra. 
These observational results suggest that Evershed flows are radial
outward flows propagating along the penumbral horizontal magnetic
fields, and their origin is rising hot gas well observed in the inner
penumbra.  
The formations of Evershed flows and penumbral fine-scale magnetic field
structures are closely related each other.

Many models have been proposed to explain penumbral fine-scale
structures and Evershed flows, two major ones are an ``uncombed''
penumbral model \citep{Solanki1993} and a ``gappy'' penumbral model
\citep{Spruit2006, Scharmer2006}. 
A nearly horizontal flux tube is embedded in more vertical background
fields in the uncombed penumbral model.  
On the basis of the this model, \citet{Schlichenmaier1998} proposed a
moving flux tube model, in which a thin flux tube carrying hot Evershed
flow rises into the background fields and moves toward the umbra.
In the gappy penumbral model, a cusp-like magnetic field structure is
formed by convection in field-free gaps just below the visible solar
surface.
Recent 3D magnetohydrodynamic (MHD) simulations with radiative transfer
are beginning to produce penumbral filaments with a systematic outflow like
observed sunspot penumbrae \citep{Heinemann2007, Rempel2009}. 
From such 3D MHD simulations, a kind of hybrid model is proposed that explains
the Evershed effect as a natural consequence of
``convection'' in strong, inclined magnetic fields \citep{Scharmer2008,
Rempel2009, Kitiashvili2009}.
Evershed flows are associated with weaker and horizontal magnetic
fields in the deep photospheric layer \citep[e.g.][]{Jurcak2007,
Borrero2008a}, but the field-free plasma has not yet been detected near
the $\tau=1$ surface in the penumbra even with spectropolarimetric
measurements at the current highest spatial resolution
\citep{Borrero2008b}. 
Moreover, \citet{Ichimoto2007b} have discovered thin inclined dark
stripes along the bright penumbral filaments suggesting overturning 
perpendicular convection within the bright penumbral filaments.

In this study, we investigate the temporal relation between the Evershed
flow, a penumbral grain, the complex magnetic field structure, and a
bright penumbral filament.
However, it is not easy to trace a penumbral filament from its birth
to death because penumbral filaments and Evershed flows are
always changing everywhere in the penumbra.
We use the dark core of a bright filament in the penumbra in order to
trace the center of the bright filament.
Such dark cores were discovered as narrow dark lanes with a width
less than 200 km in the center of bright penumbral filaments
\citep{Scharmer2002}.
The dark core is thought to be a result of a cusp shape of the $\tau=1$
surface \citep{Spruit2006, Schussler2006, Heinemann2007, Rempel2009}. 
A pile-up of hot plasma along the center of the bright penumbral
filament pushes up the the $\tau=1$ surface into the upper layer, where
the temperature is lower.  
As a result, a dark lane is observed along the top of the cusp of the
$\tau=1$ surface.
\citet{Rimmele2008} has reported that the dark cores are more clearly
observed in the line core than in the continuum, and suggested that the
dark cores form in the upper layer of the photosphere.  

The dark-cored bright filaments in the penumbra are not transient
phenomena, and they often have lifetimes on the order of 1-2 hr
\citep{Langhans2007}.  
The space-borne observations with the Solar Optical Telescope
\citep[SOT;][]{Tsuneta2008} aboard the \textit{Hinode} spacecraft
\citep{Kosugi2007} have an advantage in observing the detailed
evolution of the penumbral fine-scale structures and Evershed flows
during their lifetimes. 
We can identify a dark core with shorter wavelengths of the SOT, but the
0.3$\arcsec$ spatial resolution of the SOT spectropolarimeter is not
enough to investigate the magnetic field vector of dark cores.
The nature of dark cores is not the target of this paper.
Our goal is to investigate the nature of the magnetoconvection in the
penumbra by tracing the dark core.

\section{OBSERVATION AND DATA ANALYSIS}
\begin{figure}
\epsscale{.50}
\plotone{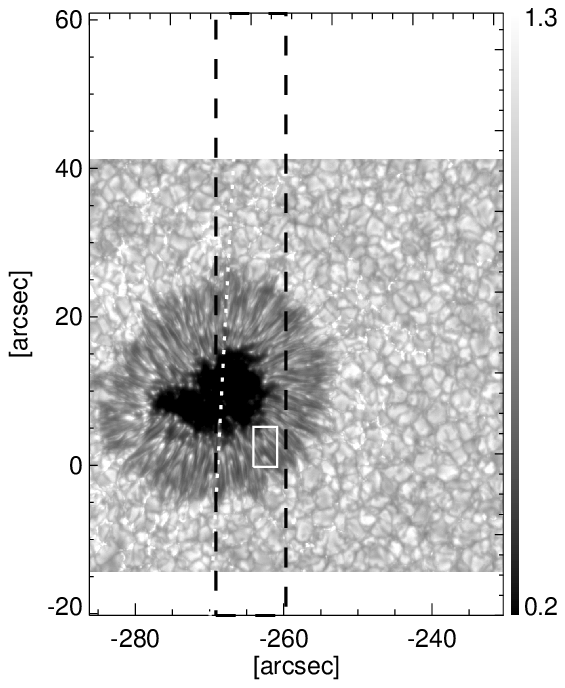}
\caption{G-band image of a sunspot in NOAA AR 10944 on 2007 February 27.
The dashed box shows the field-of-view of the \textit{Hinode} SP
and is identical to the field-of-view of Fig.~\ref{fig_FOV1}.
The dotted line is the boundary dividing the limb-side and
center-side penumbra.
The gray-scale bar shows the intensity level normalized to the mean
intensity of the quiet area outside the sunspot.
The horizontal and vertical axes represent the positions with respect to
 the disk center.} 
\label{fig_sunspot}
\end{figure}

The SOT observations of a sunspot in NOAA AR 10944 on 2007 February 27 and
calibration procedures for the observed data were described 
in \citet{Kubo2008}. 
In this Letter, we compare maps of the physical parameters obtained by
the spectropolarimeter (SP) to G-band images with the broadband filter
imager (BFI), as shown in Figure~\ref{fig_sunspot}.
The G-band images were obtained with a cadence of 2 minutes and a pixel
sampling of 0.054$\arcsec$. 
The SP repeatedly obtained spatial distributions of Stokes profiles with
a field of view of 9.5$\arcsec \times$ 82$\arcsec$ every 5.5 minutes.
The spatial sampling of SP was 0.149$\arcsec \times$ 0.160$\arcsec$.
The cadence of 5.5 minutes for the SP is insufficient to trace an
individual Evershed flow (cloud).
However, a penumbral filament is more likely to drift out of the field-of-view
during its lifetime for a higher cadence because of
the corresponding narrower field-of-view required.

The magnetic field vector and Doppler velocity in the photosphere were
derived from Stokes profiles of two Fe \small{I} lines at 630.15 and
630.25 nm by assuming a Milne-Eddington representation of the
atmospheric stratification and a two-component model atmosphere (a
magnetized atmosphere and a non-magnetized atmosphere).
We investigated the magnetic field strength, the magnetic field
inclination with respect to the line-of-sight direction, the Doppler
velocity, the continuum intensity, and the intensity at the center of
the Fe \small{I} $\lambda$ 6301.5 line.
An inclination angle of 90$\degr$ is defined as a magnetic field
perpendicular to the line-of-sight direction. 
Because the sunspot is located close to the disk center (S01, E13),
the line-of-sight direction is almost parallel with the normal direction
to the solar surface. 
This is good for investigating the interlaced vertical and horizontal
components of penumbral magnetic fields.
The reference point of the Doppler velocity in each map is the average
Doppler velocity in the umbra.  
The continuum and line center intensities are normalized to the
average continuum intensities in the quiet area outside the sunspot.

The G-band images are also normalized to average intensities in the quiet
area.
We aligned the sunspot centers in sequential G-band images in order to
remove any drifts due to the correlation tracker \citep{Shimizu2008} and
any sunspot proper motion.
The sunspot center is determined as the center-of-gravity of the G-band
intensity in the umbra.
The maps obtained by the SP were aligned to the G-band images taken at the
time closest to that of the midpoint of the SP maps, using the image
cross-correlation between the line center intensity maps and the G-band
images.

\section{RESULTS}
\begin{figure}
\epsscale{1.0}
\plotone{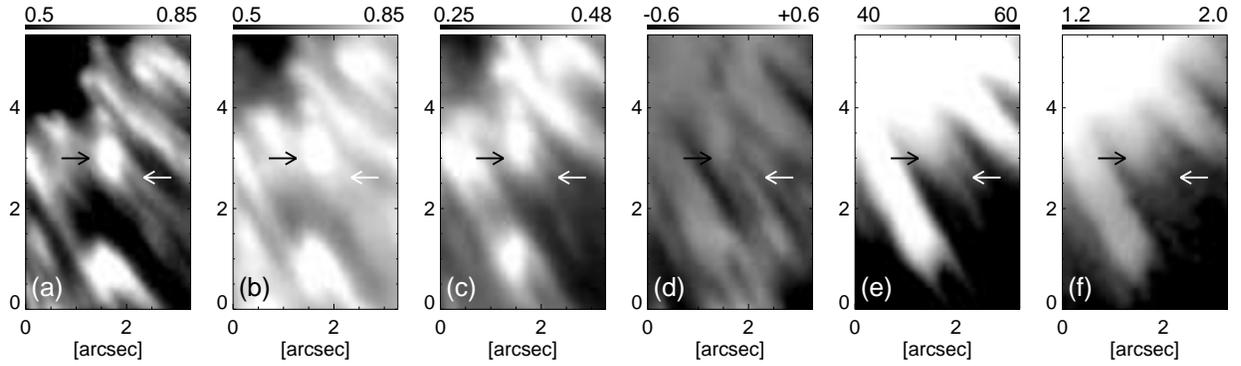}
\caption{A penumbral grain (the black arrow) and a dark core of the bright
 penumbral filament (the white arrow) in the box of Fig.~\ref{fig_sunspot}. 
(\textit{a}) G-band image, (\textit{b}) continuum intensity
 map, (\textit{c}) line center intensity map, (\textit{d}) Doppler
 velocity map, (\textit{e}) inclination map, and (\textit{f}) field
 strength map at 11:53UT on 2007 February 27.
The positive and negative Doppler velocity indicates a redshift and
 blueshift in units of km s$^{-1}$, respectively.
The magnetic field with the line-of-sight direction is represented by
 the inclination angle of 0$\degr$. 
The unit of the magnetic field strength is 10$^3$ Gauss.
} 
\label{fig_FOV1}
\end{figure}

Our target dark core (in a bright penumbral filament) is indicated by the
white arrow in Figure~\ref{fig_FOV1}\textit{a}.   
The dark core originates at a penumbral grain (the black arrow in
Fig.~\ref{fig_FOV1}\textit{a}). 
The dark-cored penumbral filament that is resolved in the G-band image
can be identified in the line center intensity map of the SP, but it is
hard to identify in the continuum intensity map
(Figs.~\ref{fig_FOV1}\textit{b} and \ref{fig_FOV1}\textit{c}).  
The difference between dark-cored penumbral filaments as observed in the
line core and continuum intensity has been already reported by
\citet{Rimmele2008}. 

Figure~\ref{fig_FOV1}\textit{d} shows that the dark core has a larger
blueshift than its lateral surroundings.
Since it is located in the center-side penumbra, this
larger blueshift corresponds to the Evershed flow (cloud).
The sunspot is located near the disk center at ($\mu>0.96$). 
Therefore the nearly horizontal Evershed flow has
a small line of sight component, but we can still
clearly identify its contribution.
The magnetic fields generally become more horizontal with increasing
distance from the sunspot center over the whole penumbra
(Fig~\ref{fig_FOV1}\textit{e}).   
The more horizontal field in the dark core is also weaker than either side as
seen in Figures~\ref{fig_FOV1}\textit{e} and \ref{fig_FOV1}\textit{f}.   
This dark core is larger than a typical width
of 90 km \citep{Scharmer2002}.
On the other hand, the properties of magnetic field vector and Doppler
velocity observed in our dark core are very similar to those in the
previous studies \citep{Bellot2005, Langhans2007, Bellot2007}. 
We therefore assume that our target, although larger than average, is
an otherwise ordinary dark-cored penumbral filament.

\begin{figure}
\epsscale{0.85}
\plotone{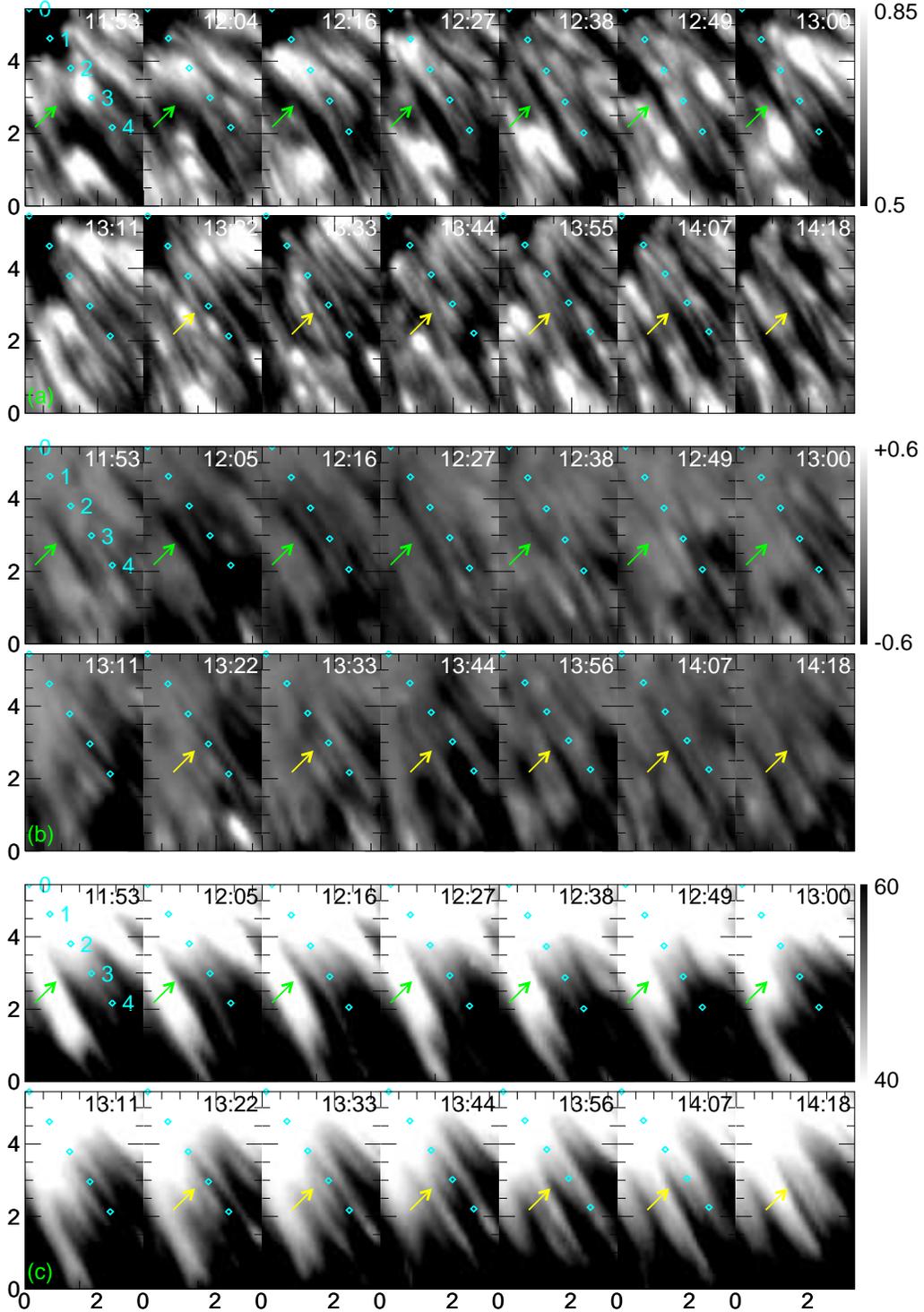}
\caption{Time series of (\textit{a}) G-band images, (\textit{b}) Doppler
 velocity maps, and (\textit{c}) inclination maps in the box of
 Fig.~\ref{fig_sunspot}.
The upper left panel in each set is identical to the corresponding signal in Figure~\ref{fig_FOV1}.
The times in panels (\textit{b}) and (\textit{c}) are for the midpoint of each SP
 map.  
The dots with intervals of 1$\arcsec$ are linearly aligned on a dark
 core in the bright penumbral filament.
The scales are in units of arcseconds.
}
\label{fig_tFOV1}
\end{figure}

We hereafter focus on the evolution of the dark-cored penumbral
filament.
The dark core is shown by the dotted lines in
Figure~\ref{fig_tFOV1}\textit{a}
which were manually determined
for each image.
The dark core elongates into the umbra in concert with the inward
motion of the penumbral grain from 11:53UT to 12:38UT. 
The dark core breaks into two parts, and the inner (umbra-side) part
shrinks into the umbra from 12:49UT to 13:11UT. 
The outer part again elongates into the umbra with another inward
migrating penumbral grain from 13:22UT to 13:44UT, and then
the dark core gradually becomes narrower and less dark from 13:44UT. 
Finally, no dark core is visible in the frame at 14:18UT. 

\begin{figure}
\epsscale{1.0}
\plotone{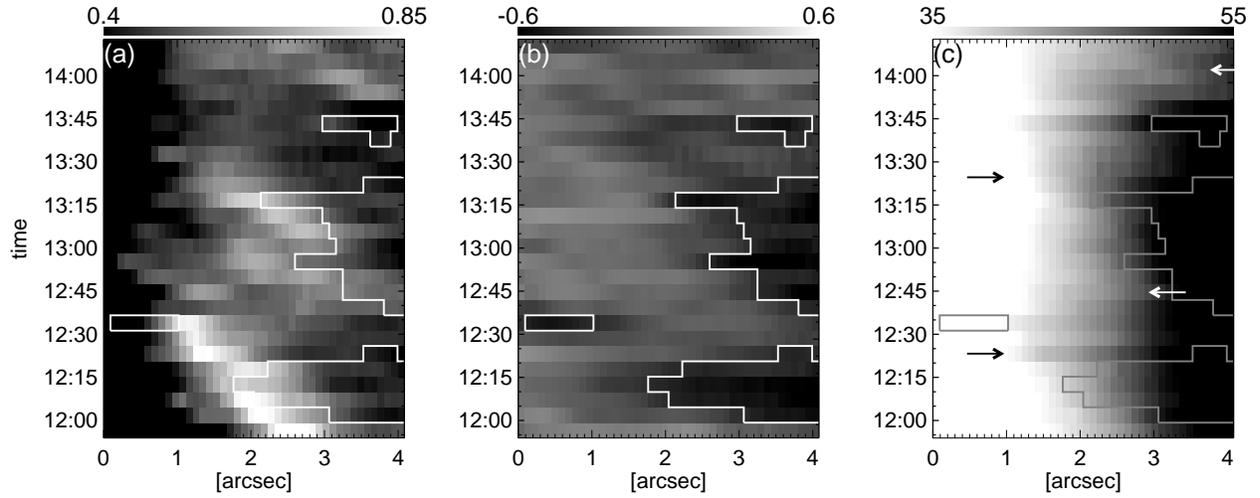}
\caption{Space vs. time plots along the dotted lines in
 Fig.~\ref{fig_tFOV1} for (\textit{a}) G-band intensity, (\textit{b})
 Doppler velocity, and (\textit{c}) magnetic field inclination from
 11:53UT to 14:12UT on 2007 February 27. 
The cadence is 5.5 minutes.
The dot at the top of each map in Fig.~\ref{fig_tFOV1} corresponds to
 0$\arcsec$ of the horizontal axis.
The contours represent the Doppler velocity of -0.4 km s$^{-1}$.
Note that the dark core is observed until 14:12UT, and not observed at
 14:18UT.  
}
\label{fig_stp1}
\end{figure}

Figure~\ref{fig_stp1}\textit{a} shows the temporal evolution of the G-band
intensity along the dotted lines in Figure~\ref{fig_tFOV1}\textit{a}.
This shows two inward moving penumbral grains,
and dark areas (dark cores) which also move into the umbra.  
The Evershed flows also originate at the penumbral grains
(Fig.~\ref{fig_stp1}\textit{b}).  
It is not observed
between the two successive penumbral grains
around 12:45UT and
it stops at 13:45 with the disappearance of the second
penumbral grain.
Small redshifts are observed along the dark-cored bright filament in
the frame at 13:56 in Figure~\ref{fig_tFOV1}\textit{b}. 
One interesting result is that the dark core is still observed
until 14:12 after the disappearance of the Evershed flow. 
Figure~\ref{fig_stp1}\textit{c} shows that the areas with more
horizontal fields move a short distance toward the umbra during the
period when Evershed flows are observed (the white arrows) in comparison
to the period without any Evershed flows (the black arrows). 
The magnetic fields at the dark core return to the same orientation as their
surroundings (more vertical) just after the penumbral grain and Evershed flow are no longer
observed along the dark core. 
Figure~\ref{fig_tFOV1}\textit{c} shows that the more horizontal fields
are observed along the dark core at 13:44 but not seen at 13:56.
The 2D distribution of magnetic field inclination is not much changed
after 13:00, and the area with vertical fields suddenly appears along the
dark core in the period between 13:44 and 13:56.
Higher time resolution is necessary to reveal the process of the
disappearance of fine-scale magnetic field structures in the penumbra.

Another dark-cored penumbral filament (the yellow arrows in
Fig~\ref{fig_tFOV1}) and a dark penumbral filament (the green arrows) are
observed in the same field-of-view. 
These dark features also have more horizontal fields than their lateral
surroundings, and have similar spatial and temporal relations with the
Evershed flows.

\section{DISCUSSIONS}
We confirm that the Evershed flow originates at penumbral grains
and is associated with a more horizontal magnetic field. 
In this paper we describe a bright penumbral filament with a
dark core that survives for 10-20 minutes beyond 
the end of
the Evershed flows. 
This means that the convective upflow along the center of the bright
penumbral filament, which produces warping of the $\tau=1$ surface (i.e., the
dark core), continues after the disappearance of the Evershed flow,
since a cooling time scale in the photosphere is much shorter than
10-20 minutes \citep{Schlichenmaier1999}.
The disappearance of the Evershed flow is locally accompanied by a
change of magnetic field inclination to more vertical.
The absence of the Evershed flow associated with convection in vertical
fields suggests that both convection and horizontal fields in the
penumbra are necessary for (or caused by) the formation of the Evershed
flow.  
The close correlation between the Evershed flow, convection, and more
horizontal fields supports recent models wherein the Evershed flows are
convective flows in the inclined magnetic field lines in the
penumbra \citep{Scharmer2008, Rempel2009, Kitiashvili2009}. 

If the Evershed flow only caused a pile-up of hot plasma along
the bright filament, the dark core should disappear within a cooling
time scale after the end of the Evershed flow. 
However, the dark core is continuously observed without
the Evershed flow for a period longer than a cooling time scale in the
photosphere. 
This suggests that a heat source, in addition to the heat transfer by
the Evershed flows, is necessary to form a bright filament with a dark core.
It is believed that the Evershed flows are radially outward flows along
the horizontal fields from the inner penumbra (penumbral grains) to the
outer penumbra.  
They may carry the energy through the penumbra in the horizontal
direction rather than the vertical direction. 
Although it may be hard to identify the Evershed flows from convection
in the penumbra, one possible source of the heat for the bright
penumbral filament is vertical convective motions locally formed along
the bright filament rather than the larger scale heat transfer due 
to the Evershed flows.  

A major remaining issue is the vertical stratifications in
the penumbral fine-scale structures.
After the Evershed flow disappears in the observed layer, it
may remain in the deeper photosphere or below
the photosphere.
Recently, signatures of convection (i.e. downward and upward motions) in
umbral dots \citep{Ortiz2010} and light bridges \citep{Rouppe2010} are
reported only in the deep photosphere from a bisector analysis of the Fe
I 630 nm lines taken with the CRISP at the Swedish 1 m Solar Telescope. 
However, the signature of convective downflows is not detected from the same
analysis of Fe I 709 nm line taken with the Littrow spectrograph at the same
telescope \citep{Bellot2010}.  
It is still unknown how the Evershed flows in the deeper layer affect the
formation of the convection in the observed layer and heating of bright
filaments.
Moreover, the evolution of the vertical stratifications is essential for
understanding the formation and decay of the penumbral complex magnetic
field structures.  
We have observed redshifts along the dark core when the more horizontal 
magnetic field disappears from the dark-cored penumbral filament, but
they are very small. 
It is not clear if the observed redshifts correspond to the
downward motion of the magnetic field lines because a flow along the
field lines also can produce such a small redshift. 
If the more horizontal fields have either upward motion or downward
motion during their disappearance, the vertical stratifications of the
magnetic field vector should give us information about their motion.

\acknowledgments
We would like to thank J. M. Borrero, M. Rempel, R. Schlichenmaier,
Y. Katsukawa, A. Lagg, J. Jur{\v c}{\'a}k, T. Magara, and S. Tsuneta for
fruitful comments.  
Hinode is a Japanese mission developed and launched by ISAS/JAXA, with
NAOJ as domestic partner and NASA and STFC (UK) as international
partners. It is operated by these agencies in co-operation with ESA and
NSC Norway. 
This work was partly carried out at the NAOJ Hinode Science Center,
which is supported by the Grant-in-Aid for Creative Scientific Research
``The Basic Study of Space Weather Prediction'' from MEXT, Japan (Head
Investigator: K. Shibata), generous donations from Sun Microsystems, and
NAOJ internal funding.

\begin{thebibliography}{}
\bibitem[Bellot Rubio et al.(2004)]{Bellot2004} Bellot Rubio, L.~R., Balthasar, H., \& Collados, M.\ 2004, \aap, 427, 319 
\bibitem[Bellot Rubio et al.(2005)]{Bellot2005} Bellot Rubio, L.~R., Langhans, K., \& Schlichenmaier, R.\ 2005, \aap, 443, L7 
\bibitem[Bellot Rubio et al.(2007)]{Bellot2007} Bellot Rubio, L.~R., et al.\ 2007, \apjl, 668, L91 
\bibitem[Bellot Rubio et al.(2010)]{Bellot2010} Bellot Rubio, L.~R., Schlichenmaier, R., \& Langhans, K.\ 2010, \apj, 725, 11 
\bibitem[Borrero(2007)]{Borrero2007} Borrero, J.~M.\ 2007, \aap, 471, 967 
\bibitem[Borrero et al.(2008)]{Borrero2008a} Borrero, J.~M., Lites, B.~W., \& Solanki, S.~K.\ 2008, \aap, 481, L13 
\bibitem[Borrero \& Solanki(2008)]{Borrero2008b} Borrero, J.~M., \& Solanki, S.~K.\ 2008, \apj, 687, 668 

\bibitem[Degenhardt \& Wiehr(1991)]{Degenhardt1991} Degenhardt, D., \& Wiehr, E.\ 1991, \aap, 252, 821 


\bibitem[Heinemann et al.(2007)]{Heinemann2007} Heinemann, T., Nordlund, {\AA}., Scharmer, G.~B., \& Spruit, H.~C.\ 2007, \apj, 669, 1390 

\bibitem[Ichimoto et al.(2007a)]{Ichimoto2007a} Ichimoto, K., et al.\ 2007, \pasj, 59, 593 
\bibitem[Ichimoto et al.(2007b)]{Ichimoto2007b} Ichimoto, K., et al.\ 2007, Science, 318, 1597 

\bibitem[Jurc{\'a}k et al.(2007)]{Jurcak2007} Jurc{\'a}k, J., et al.\ 2007, \pasj, 59, 601 

\bibitem[Kitiashvili et al.(2009)]{Kitiashvili2009} Kitiashvili, I.~N., 
Kosovichev, A.~G., Wray, A.~A., \& Mansour, N.~N.\ 2009, \apjl, 700, L178 
\bibitem[Kosugi et al.(2007)]{Kosugi2007} Kosugi, T., et al.\ 2007, \solphys, 243, 3 
\bibitem[Kubo et al.(2008)]{Kubo2008} Kubo, M., et al.\ 2008, \apj, 681, 1677 

\bibitem[Langhans et al.(2005)]{Langhans2005} Langhans, K., Scharmer, G.~B., Kiselman, D., L{\"o}fdahl, M.~G., \& Berger, T.~E.\ 2005, \aap, 436, 1087 
\bibitem[Langhans et al.(2007)]{Langhans2007} Langhans, K., Scharmer, G.~B., Kiselman, D., \& L{\"o}fdahl, M.~G.\ 2007, \aap, 464, 763 
\bibitem[Lites et al.(1993)]{Lites1993} Lites, B.~W., Elmore, D.~F., Seagraves, P., \& Skumanich, A.~P.\ 1993, \apj, 418, 928 

\bibitem[Ortiz et al.(2010)]{Ortiz2010} Ortiz, A., Bellot Rubio, L.~R., \& Rouppe van der Voort, L.\ 2010, \apj, 713, 1282 


\bibitem[Rempel et al.(2009)]{Rempel2009} Rempel, M., Sch{\"u}ssler, M., \& Kn{\"o}lker, M.\ 2009, \apj, 691, 640 
\bibitem[Rimmele \& Marino(2006)]{Rimmele2006} Rimmele, T., \& Marino, J.\ 2006, \apj, 646, 593 
\bibitem[Rimmele(2008)]{Rimmele2008} Rimmele, T.\ 2008, \apj, 672, 684 
\bibitem[Rouppe van der Voort et al.(2010)]{Rouppe2010} Rouppe van der Voort, L., Bellot Rubio, L.~R., \& Ortiz, A.\ 2010, \apjl, 718, L78 


\bibitem[Scharmer et al.(2002)]{Scharmer2002} Scharmer, G.~B., Gudiksen, B.~V., Kiselman, D., L{\"o}fdahl, M.~G., \& Rouppe van der Voort, L.~H.~M.\ 2002, \nat, 420, 151 
\bibitem[Scharmer \& Spruit(2006)]{Scharmer2006} Scharmer, G.~B., \& Spruit, H.~C.\ 2006, \aap, 460, 605 
\bibitem[Scharmer et al.(2008)]{Scharmer2008} Scharmer, G.~B., Nordlund, {\AA}., \& Heinemann, T.\ 2008, \apjl, 677, L149 
\bibitem[Schmidt et al.(1992)]{Schmidt1992} Schmidt, W., Hofmann, A., Balthasar, H., Tarbell, T.~D., \& Frank, Z.~A.\ 1992, \aap, 264, L27 
\bibitem[Schlichenmaier et al.(1998)]{Schlichenmaier1998} Schlichenmaier, R., Jahn, K., \& Schmidt, H.~U.\ 1998, \aap, 337, 897 
\bibitem[Schlichenmaier et al.(1999)]{Schlichenmaier1999} Schlichenmaier, R., Bruls, J.~H.~M.~J., \& Sch{\"u}ssler, M.\ 1999, \aap, 349, 961 
\bibitem[Sch{\"u}ssler \& V{\"o}gler(2006)]{Schussler2006} Sch{\"u}ssler, M., \& V{\"o}gler, A.\ 2006, \apjl, 641, L73 
\bibitem[Shimizu et al.(2008)]{Shimizu2008} Shimizu, T., et al.\ 2008, \solphys, 249, 221 
\bibitem[Sobotka et al.(1999)]{Sobotka1999} Sobotka, M., Brandt, P.~., \& Simon, G.~W.\ 1999, \aap, 348, 621 
\bibitem[Solanki \& Montavon(1993)]{Solanki1993} Solanki, S.~K., \& Montavon, C.~A.~P.\ 1993, \aap, 275, 283 
\bibitem[Spruit \& Scharmer(2006)]{Spruit2006} Spruit, H.~C., \& Scharmer, G.~B.\ 2006, \aap, 447, 343 
\bibitem[S{\"u}tterlin et al.(2004)]{Sutterlin2004} S{\"u}tterlin, P., Bellot Rubio, L.~R., \& Schlichenmaier, R.\ 2004, \aap, 424, 1049 

\bibitem[Title et al.(1993)]{Title1993} Title, A.~M., Frank, Z.~A., Shine, R.~A., Tarbell, T.~D., Topka, K.~P., Scharmer, G., \& Schmidt, W.\ 1993, \apj, 403, 780 
\bibitem[Tsuneta et al.(2008)]{Tsuneta2008} Tsuneta, S., et al.\ 2008, \solphys, 249, 167 
\end{thebibliography}
\end{document}